\documentclass[lettersize,journal]{IEEEtran}
\usepackage{amsmath,amsfonts}
\usepackage{algorithmic}
\usepackage{algorithm}
\usepackage{array}
\usepackage[caption=false,font=normalsize,labelfont=sf,textfont=sf]{subfig}
\usepackage{textcomp}
\usepackage{stfloats}
\usepackage{url}
\usepackage{verbatim}
\usepackage{graphicx}
\usepackage{cite}
\usepackage{xurl}
\usepackage{float}
\usepackage{booktabs}
\usepackage{multirow}
\usepackage{graphicx}
\usepackage{amsmath}
\usepackage{makecell}

\begin{document}

\title{DiT-Flow: Speech Enhancement Robust to Multiple Distortions based on Flow Matching in Latent Space and Diffusion Transformers}

\author{Tianyu Cao, Helin Wang, Ari Frummer, Yuval Sieradzki, Adi Arbel, Laureano Moro Velazquez~\IEEEmembership{Member,~IEEE}, \\
Jes\'us Villalba,~\IEEEmembership{Member,~IEEE}, Oren Gal,~\IEEEmembership{Member,~IEEE}, Thomas Thebaud,~\IEEEmembership{Member,~IEEE}, \\ Najim Dehak,~\IEEEmembership{Senior~Member,~IEEE}

\thanks{
Tianyu Cao, Helin Wang, Ari Frummer, Laureano Moro Velazquez, Jes\'us Villalba, Thomas Thebaud and Najim Dehak are with the Johns
Hopkins University (email: $\{$tcao7, hwang258, afrumme1, laureano, jvillal7, tthebau1, ndehak3$\}$@jhu.edu).}

\thanks{
Yuval Sieradzki and Adi Arbel are with the Technion Israel Institute of Technology (email: $\{$syuvsier, adi.arbel$\}$@campus.technion.ac.il). Oren Gal is with the University of Haifa, (email: orengal@univ.haifa.ac.il).
}}

\maketitle

\begin{abstract}

Recent advances in generative models, such as diffusion and flow matching, have shown strong performance in audio tasks. However, speech enhancement (SE) models are typically trained on limited datasets and evaluated under narrow conditions, limiting real-world applicability. To address this, we propose DiT-Flow, a flow matching-based SE framework built on the latent Diffusion Transformer (DiT) backbone and trained for robustness across diverse distortions, including noise, reverberation, and compression. DiT-Flow operates on compact variational auto-encoders (VAEs)-derived latent features. We validated our approach on StillSonicSet, a synthetic yet acoustically realistic dataset composed of LibriSpeech, FSD50K, FMA, and 90 Matterport3D scenes. Experiments show that DiT-Flow consistently outperforms state-of-the-art generative SE models, demonstrating the effectiveness of flow matching in multi-condition speech enhancement. Despite ongoing efforts to expand synthetic data realism, a persistent bottleneck in SE is the inevitable mismatch between training and deployment conditions. By integrating LoRA with the MoE framework, we achieve both parameter-efficient and high-performance training for DiT-Flow robust to multiple distortions with using 4.9\% percentage of the total parameters to obtain a better performance on five unseen distortions. 
\end{abstract}

\begin{IEEEkeywords}
Speech Enhancement, Generative Model, Synthetic Data, StillSonicSet, Domain Adaptation
\end{IEEEkeywords}

\section{Introduction}

Speech enhancement (SE) aims to reconstruct clean speech from signals corrupted by environmental noise~\cite{loizou2007speech}. Classical methods relied on explicit statistical models of speech and noise distributions~\cite{gerkmann2018spectral, kim2022improved, ephraim1984speech}, whereas contemporary research predominantly employs deep neural networks (DNNs) to estimate either the clean waveform directly or a multiplicative mask applied to noisy inputs~\cite{song2022speech, hu2020dccrn, tan2019learning, tan2018gated}. Recently, generative approaches, which learn the underlying distribution of clean speech signals, have become powerful alternatives for SE~\cite{baby2019sergan, nugraha2020flow, bie2022unsupervised, richter2023speech, lemercier2023storm, lay2024single}. In particular, score-based generative models, also known as diffusion models, formulated as stochastic differential equations (SDEs), have recently achieved impressive performance~\cite{richter2023speech, lemercier2023storm}. These models reconstruct clean speech by numerically solving the corresponding reverse SDE, a procedure that involves repeatedly estimating the score function. Therefore, diffusion-based methods are computationally intensive and exhibit high latency, which constrains their feasibility for real-time applications. Recently, flow matching (FM), introduced in~\cite{lipman2022flow}, has recently emerged as a promising alternative to diffusion-based methods for training continuous normalizing flows (CNFs)~\cite{tong2023improving}, which is now applied to different tasks, e.g., speech‐processing applications~\cite{liu2023generative}, and audio–visual speech enhancement~\cite{jung2024flowavse}. Unlike diffusion models, which rely on successive stochastic denoising steps for inference, FM learns a deterministic, time-varying velocity field that enables a single, smoothly guided transformation from Gaussian noise to the target data distribution. Moreover, recent studies have shown that existing diffusion models can be optimized using the FM objective instead of standard score matching, resulting in substantially faster sampling during inference. Although early attempts has applied flow matching to speech enhancement \cite{wang2025flowse}, the method has not been fully explored in latent space, which potentially fulfill the task at larger scale but lower computation cost and time.

A systematic evaluation of speech enhancement models under far-field conditions requires large-scale, diverse datasets that capture real-world acoustic variability. However, existing real-world datasets are often limited in size and diversity, constraining both training and robust evaluation. Synthetic datasets, while scalable, typically fall short in acoustic realism. For instance, they often rely on simplified room impulse responses (RIRs) generated using the image source method, which assumes idealized conditions such as empty, box-shaped rooms. This abstraction fails to account for critical factors present in realistic environments, including the occlusion of obstacles, where sound paths are blocked or diffracted by furniture or human bodies, as well as complex room geometries and heterogeneous surface materials that influence sound propagation in nuanced ways. Consequently, there is a pressing need for more representative datasets or advanced simulation techniques that better reflect the acoustic complexity of real-world far-field scenarios. Recently, a synthetic toolkit, SonicSim was proposed, which enables the synthesis of acoustically diverse datasets~\cite{li2024sonicsim}. A moving-sound source benchmark dataset named SonicSet was constructed by SonicSim and compared with existing synthetic datasets, which show that models trained on SonicSet achieve markedly stronger generalization to real-world conditions than those trained on other synthetic corpora~\cite{li2024sonicsim}.

However, in some scenarios, such as meetings, teleconferencing, or classroom settings, the speaker or sound source typically remains stationary. While SonicSim and SonicSet have advanced the generation of acoustically diverse data for moving sources, there remains a notable gap in synthetic datasets tailored to stationary sound sources under complex real-world acoustic conditions. These static-source environments are equally important for practical speech enhancement and separation tasks, yet current synthetic datasets still fall short in capturing the rich spatial and material diversity encountered in such scenarios. Besides, real-time speech communication systems, such as VoIP, teleconferencing, and mobile calls, rely heavily on low-delay audio codecs to compress speech signals under stringent bandwidth constraints. Among these, the Opus codec has become a widely adopted standard due to its flexibility and low-latency properties. Opus is a royalty-free, open audio codec standardized by the Internet Engineering Task Force (IETF), designed for real-time interactive applications such as voice over IP (VoIP), video conferencing, and streaming. However, Opus introduces significant compression artifacts, including quantization noise, spectral smearing, and loss of fine-grained detail~\cite{britanak2018audio}. These artifacts can significantly degrade perceived audio quality, especially for expressive, tonal, or music-rich speech content. Besides, speech signals captured in natural environments are frequently contaminated by ambient noise, reverberation, and recording artifacts. Although deep learning-based speech enhancement models have achieved impressive denoising results~\cite{lemercier2023storm}, they typically assume access to uncompressed inputs with additive environmental noise. This assumption limits their applicability in real-world pipelines where speech is often already compressed and corrupted.

Despite ongoing efforts to expand synthetic data realism, a persistent bottleneck in SE is the inevitable mismatch between training and deployment conditions. As a result, SE models that perform well on matched test sets often degrade noticeably under domain shift, motivating research on \emph{data adaptation} methods that can adjust models to new acoustic conditions with limited in-domain evidence. Many recent studies address this issue through domain adaptation and test-time adaptation. One of the practical settings is \emph{few-shot} adaptation, where a small amount of in-domain data is available, either paired (noisy, clean) samples, weak supervision, or a tiny support set from a new noise, speaker or environment. Meta-learning has been explored to enable fast adaptation with only a few examples, e.g., Meta-SE for few-shot noise adaptation \cite{zhou2021meta} and one-shot speaker-adaptive SE via meta-learning \cite{yu2022ossem}. Such approaches reflect realistic deployment constraints. That is, collecting large-scale, fully supervised in-domain datasets is often infeasible, but obtaining a handful of short recordings from a target device, room, or codec setting is comparatively easy. However, na\"ive fine-tuning, even with few-shot data, can be computationally expensive for modern high-capacity SE systems and may overfit or catastrophically forget previously learned generalization. This motivates \emph{parameter-efficient and fine-tuning} (PEFT) strategies, which freeze the backbone model and update only a small set of trainable parameters, such as adapter modules \cite{houlsby2019parameter}, low-rank updates (LoRA) \cite{hu2022lora}, prefix-style continuous prompts \cite{li2021prefix}, or even sparse bias-only updates \cite{zaken2021bitfit}. However, a single adapted parameter set (e.g., one LoRA) may be insufficient when the target distribution itself is heterogeneous or non-stationary. This motivates \emph{Mixture-of-Experts} (MoE) modeling, where conditional computation activates different expert subnetworks for different inputs, substantially increasing representational capacity without proportionally increasing inference cost \cite{fedus2022switch}. Combining the strengths of LoRA and MoE yields a particularly appealing adaptation mechanism, \emph{MoELoRA}, where each expert corresponds to a lightweight LoRA module (or a small set of LoRA modules) while the backbone remains frozen, which suggests a promising direction for SE under real-world mismatch. Rather than relying solely on a single globally trained model or a single adapted LoRA, MoELoRA can maintain a compact library of specialists and dynamically activate the most relevant ones for the current distortion conditions, making few-shot adaptation and continual deployment more feasible. Prior SE work has explored expert-style specialization and routing for personalization and test-time specificity, demonstrating that selectively using specialized modules can improve robustness under changing speaker and environment characteristics \cite{sivaraman2021zeros,kim2021testtime}. 

Although parameter-efficient adaptation offers a promising path toward robust SE in realistic pipelines, where training datasets cannot fully anticipate the distortions encountered at test time, there is a lack of exploration of the combination of both LoRA and MoE together for data adaptation and parameter-efficiency in speech enhancement systems. The key contributions of our work are summarized as follows:

\begin{enumerate}
    \item We propose a DiT-Flow model, a flow matching-based SE framework built on the latent Diffusion Transformer (DiT) backbone, that is robust across multiple common and compound distortions, including noise, reverberation, and compression artifacts.
    \item We introduce StillSonicSet, a newly constructed synthetic dataset with acoustically realistic conditions by incorporating complex room geometries, varied surface materials, and natural occlusions such as furniture and architectural structures.
    \item We conduct comprehensive experiments by training DiT-Flow on our StillSonicSet and validate its robustness against multiple distortions, showing that DiT-Flow consistently outperforms state-of-the-art generative SE models, thereby demonstrating the effectiveness of flow matching in multi-condition speech enhancement.
    \item We first apply LoRA with the MoE framework to a generative speech enhancement system to adapt multiple distortions, achieving both parameter-efficient and high-performance training.
\end{enumerate}

\section{Background}

\subsection{Flow Matching for Generative Modeling}

Flow Matching, proposed in~\cite{lipman2022flow}, trains Continuous Normalizing Flows (CNF)~\cite{chen2018neural} on Euclidean space that sidesteps likelihood computation during training. This method has already been successfully applied in different speech tasks, e.g., speech separation~\cite{yuan2025flowseplanguagequeriedsoundseparation}, text-to-speech (TTS)~\cite{mehta2024matchattsfastttsarchitecture}, and audio–visual speech enhancement~\cite{jung2024flowavse}. 

CNF models an invertible mapping $\phi_t:[0,1] \times \mathbb{R}^d \rightarrow \mathbb{R}^d$ to transform from a simple space with a known distribution $p\left(x_0\right)$, e.g., $\mathcal{N}(\mathbf{0}, \mathbf{I})$, to another space distributed by an unknown distribution denoted as $q\left(x_1\right)$ for which only samples are available. The invertible mapping is defined as

\begin{equation}
\label{eq:ode}
\frac{d}{d t} \phi_t\left(x_0\right)=v_t\left(\phi_t\left(x_0\right)\right), \quad \phi_0\left(x_0\right)=x_0 \;,
\end{equation}

where $v_t:[0,1] \times \mathbb{R}^d \rightarrow \mathbb{R}^d$ is a time-dependent velocity (vector) field and $d$ denotes the data dimension. Given the flow, the density of the intermediate state $x_t=\phi_t\left(x_0\right), p_t\left(x_t\right)$ is given by

\begin{equation}
p_t\left(x_t\right)=p_0\left(\phi_t^{-1}\left(x_t\right)\right) \operatorname{det}\left[\frac{\partial \phi_t^{-1}\left(x_t\right)}{\partial x_t}\right]\;.
\end{equation}

Training a CNF aims to learn a velocity field $v_t$ (or equivalently $\phi_t$ ) such that $p_0\left(x_0\right)=p\left(x_0\right)$ and the terminal density $p_1\left(x_1\right)$ closely matches $q\left(x_1\right)$. In practice, $v_t$ is parameterised by a neural network $v_\theta(x, t)$. In~\cite{lipman2022flow}, the conditional flow-matching (CFM) loss is introduced to consider the tractable conditional density $p_t\left(x_t \mid x_1\right)$ and the associated conditional vector field $v_t\left(x_t \mid x_1\right)$ rather than $p_t\left(x_t\right)$ and $v_t\left(x_t\right)$. The CFM loss given by

\begin{equation}
\mathcal{L}_{C F M}(\theta):=\mathbf{E}_{t, x_1, p_t\left(x_t \mid x_1\right)}\left\|v_\theta\left(x_t, t\right)-v_t\left(x_t \mid x_1\right)\right\|^2\;.
\end{equation}

\subsection{Mixture-of-LoRA Experts}
\label{sec:Mixture-of-LoRA Experts}
Recent advancements in LLMs have resulted in the development of efficient methods that enhance scalability and generalization. In this work, we investigate the fusion of MoE and LoRA, i.e., Mixture-of-LoRA Experts, for potential improvements in speech enhancement.

\subsubsection{Low-Rank Adapters.}

Initially, Low-Rank Adaptation (LoRA) is a parameter-efficient fine-tuning strategy designed to adapt large pre-trained neural networks \cite{hu2022lora}, especially transformer-based foundation models without updating the full set of backbone weights. Its core premise is that, for many downstream tasks, the effective change required to a pre-trained weight matrix lies in a low-dimensional subspace. LoRA operationalizes this premise by constraining the fine-tuning-induced weight update to be low-rank, thereby reducing the number of trainable parameters.

Consider a linear transformation in a pre-trained model with weight matrix $\mathbf{W}_0 \in \mathbb{R}^{d \times \ell}$. Standard finetuning learns an updated matrix

\begin{equation}
\mathbf{W}=\mathbf{W}_0+\Delta \mathbf{W},
\end{equation}

where $\Delta \mathbf{W}$ is a dense matrix of the same shape as $\mathbf{W}_0$. This dense update is expensive when $d$ and $\ell$ are large and must be repeated for each downstream task or domain.

LoRA replaces the unconstrained dense update with a structured low-rank factorization:

\begin{equation}
\mathbf{\Delta} \mathbf{W} \approx \mathbf{B A},
\end{equation}

where $\mathbf{B} \in \mathbb{R}^{d \times r}$ and $\mathbf{A} \in \mathbb{R}^{r \times \ell}$, and the rank $r$ is chosen such that $r \ll \min \{d, \ell\}$. The model's pretrained parameters $\mathbf{W}_0$ remain frozen; only $\mathbf{A}$ and $\mathbf{B}$ are optimized. To control the magnitude of the injected adaptation and stabilize training across ranks, LoRA commonly introduces a scaling term:

\begin{equation}
\mathbf{W}=\mathbf{W}_0+\frac{\alpha}{r} \mathbf{B} \mathbf{A},
\end{equation}

where $\alpha$ is a tunable scaling factor. This parameterization ensures that changing $r$ does not automatically change the typical update scale, making comparisons across ranks more meaningful and improving optimization behavior.

Forward computation and module structure
In a linear layer, for an input $\mathbf{x}$, the adapted output becomes

\begin{equation}
\mathbf{h}=\mathbf{W} \mathbf{x}=\mathbf{W}_0 \mathbf{x}+\Delta \mathbf{W} \mathbf{x}=\mathbf{W}_0 \mathbf{x}+\frac{\alpha}{r} \mathbf{B} \mathbf{A} \mathbf{x} .
\end{equation}

\subsubsection{Mixture-of-Experts.}

Mixture-of-Experts (MoE) is a modular modeling paradigm that increases capacity by utilizing multiple specialized sub-networks, i.e., experts, under a learned routing mechanism, while maintaining a low number of parameters for prediction and training. Introduced in \cite{jacobs1991adaptive}, MoE has since been widely adopted in speech processing \cite{you2021speechmoe}, natural language understanding \cite{fedus2022switch}, and other application domains. Let an MoE layer contain $N$ experts $\left\{E_i(\mathbf{x})\right\}_{i=1}^N$, where each expert is a learnable transformation (commonly instantiated with feed-forward networks in practice). Given an input representation $\mathbf{x} \in \mathbb{R}^d$, a gating (routing) network produces a nonnegative weight for each expert and uses these weights to coordinate expert contributions. A standard parameterization computes gating scores via a trainable matrix $\mathbf{W}_g$ and normalizes them with a Softmax.

\begin{equation}
G_i(\mathbf{x})=\operatorname{Softmax}\left(\mathbf{W}_g \mathbf{x}+\epsilon\right)_i,
\end{equation}

where $\epsilon \sim \mathcal{N}\left(\mu, \sigma^2 I\right)$ is Gaussian noise with learnable mean $\mu$ and variance $\sigma^2$. To obtain computational sparsity, sparse MoE adopts a Top- $k$ routing rule that selects only the $k$ most highly weighted experts for each $\mathbf{x}$, with $k \ll N$. Denote the selected index set by

\begin{equation}
\mathcal{S}(\mathbf{x})=\operatorname{TopK}\left\{G_i(\mathbf{x})\right\} .
\end{equation}

The MoE output is then computed as the weighted combination of the selected experts:

\begin{equation}
\mathbf{y}=\sum_{i \in \mathcal{S}(\mathbf{x})} G_i(\mathbf{x}) E_i(\mathbf{x}) .
\end{equation}

When $k=N$, the same formulation reduces to a dense variant in which all experts contribute. In contrast, sparse routing ensures that only a limited number of experts participate in both forward computation and gradient updates for each input. Consequently, MoE can scale overall capacity by adding experts, while maintaining roughly constant computation per example by keeping $k$ fixed.

\subsubsection{Mixture of LoRA Experts}

Mixture-of-LoRA-Experts combines the parameter-efficient adaptation of LoRA with the input-conditioned specialization of Mixture-of-Experts (MoE). The core idea is to treat multiple LoRA branches as a set of experts and to use a routing (gating) network to dynamically weight and select which LoRA experts contribute to the update for a given input. The goals of this design are to simultaneously preserve the frozen backbone’s general knowledge, increase adaptation capacity through multiple specialized low-rank updates, and control computation by activating only a subset of experts when sparse routing is used. Compared with LoRA, the single low-rank update $\Delta \mathrm{W}$ is replaced by a mixture of multiple low-rank updates, each associated with a LoRA expert. Specifically, each expert $i$ has its own low-rank pair ( $\mathrm{A}_i, \mathrm{~B}_i$ ).The routing network produces weights $G_i(\mathbf{x})$ and potentially a sparse selected set $\mathcal{S}(\mathbf{x})$. The fused output takes the form as

\begin{equation}
\mathbf{h}=\mathbf{W}_0 \mathbf{x}+\sum_{i \in \mathcal{S}(\mathbf{x})} G_i(\mathbf{x})\left(\mathrm{A}_i \mathrm{~B}_i \mathbf{x}\right),
\end{equation}

where $\mathcal{S}(\mathbf{x})$ denotes the selected experts for input $\mathbf{x}$ (e.g., via Top- $k$ routing), and the term $\mathrm{A}_i \mathrm{~B}_i$ plays the same role as the LoRA low-rank update in the single-expert case, but now specialized per expert.

\section{The Still-SonicSet Dataset}
\label{sec:materials}

The recently introduced synthetic toolkit, SonicSim, supports multi-scale configuration, including scene-level, microphone-level, and source-level adjustments, which enables the synthesis of acoustically diverse datasets~\cite{li2024sonicsim}. Utilizing SonicSim, a moving sound source benchmark dataset named SonicSet was constructed. It integrates audio samples from LibriSpeech, Freesound Dataset 50k (FSD50K), and the Free Music Archive (FMA), alongside 90 distinct environments from Matterport3D, to support systematic evaluation of speech separation and enhancement models~\cite{li2024sonicsim}. 

\begin{figure}
  \centering
\includegraphics[width=8cm]{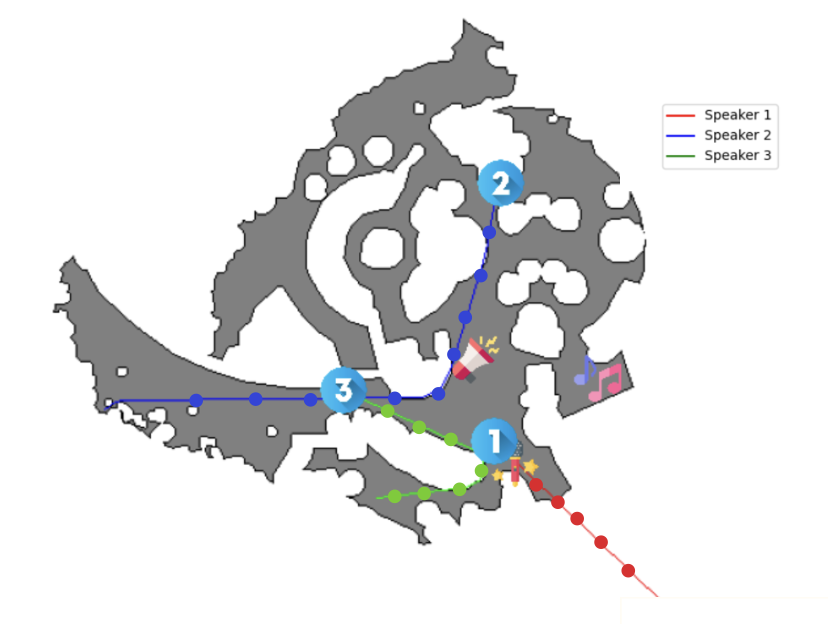}
\caption{The procedure to generate StillSonicSet. In each scene, three moving RIRs for each speaker in the original SonicSet were discretized to obtain RIR at some fixed places (circles).} 
\label{fig:rir}
\end{figure}

To generate a dataset that closely resembles real-world meeting scenarios, characterized by limited speaker movement, we leveraged the room impulse responses (RIRs) provided in SonicSet~\cite{li2024sonicsim}. These RIRs were simulated using 90 distinct environments from the Matterport3D dataset. We validated the reduced acoustic gap compared to existing synthetic datasets and demonstrated better generalization to real-world conditions.
As illustrated in Figure~\ref{fig:rir}, for each scene in SonicSet, the moving RIRs were discretized to obtain responses at fixed positions (represented as circles in the figure). The positions of sound sources, noise sources, and microphones were randomly generated within each room. Specifically, the initial position of each speech source and the position of noise sources were placed within a 1–8 meter radius of the microphone.
To simulate stationary speakers, as typically observed in meeting scenarios, speech utterances were convolved with the RIR at a single fixed position. Importantly, the volume was not normalized across different positions; instead, it naturally varied with the distance from the microphone, better reflecting real-world acoustic behavior.

We used the RIRs described above to construct StillSonicSet, a variant of SonicSet designed to simulate stationary speakers in realistic acoustic environments. For each scene in the dataset, three speech utterances from different speakers were randomly selected from LibriSpeech. These utterances were downsampled to 8 kHz and convolved with three distinct RIRs to simulate corresponding far-field speech signals.
To model environmental audio within the same scene, background noise and music samples were randomly selected from the Freesound Dataset 50k (FSD50K) and the Free Music Archive (FMA). These were convolved with other RIRs from the same acoustic environment as the speech to ensure consistent spatial characteristics.
The key properties of StillSonicSet are summarized in Table~\ref{tab:1}.

\begin{table}
\caption{StillSonicSet and its characteristics. “Speaker” refers to the number of unique speakers. “Room Style” denotes the number of different rooms. “Scenario” stands for the total number of conversations between two speakers in one room, and “Duration” represents the total duration of the dataset in hours.}

\begin{center}
\begin{tabular}{|c|c|c|c|c|}
\hline & \textbf{Speaker} & \textbf{Room Style} & \textbf{Scenario} & \textbf{Duration (h)} \\
\hline \textbf{Training} & 921 & 60 & 34672 & 363.6 \\
\hline \textbf{Validation} & 40 & 19 & 901 & 8 \\
\hline \textbf{Test} & 40 & 9 & 873 & 8 \\
\hline
\end{tabular}
\label{tab:1}
\end{center}
\vspace{-2mm}
\end{table}

To simulate realistic low-bitrate audio transmission scenarios, including teleconferencing or mobile streaming, we applied audio compression using the Opus codec via the opuslib Python package\footnote{\url{https://pypi.org/project/opuslib/}}. The compression configuration was designed to prioritize low-bitrate encoding while preserving audio intelligibility under constrained bandwidth. The following describes the compression operations in detail. The original audio, sampled at 8 kHz, was represented at 16-bit precision per sample. Each Opus encoding frame covered a 20-ms window, balancing compression efficiency with temporal resolution and latency. The encoder was configured at the maximum complexity level of 10, with 10 representing the highest complexity. The target bitrate was randomly sampled within the range of 30 kbps to 40 kbps, to reflect realistic streaming or transmission conditions, where available bandwidth may fluctuate across sessions or users.

\begin{figure}
  \centering
\includegraphics[width=8cm]{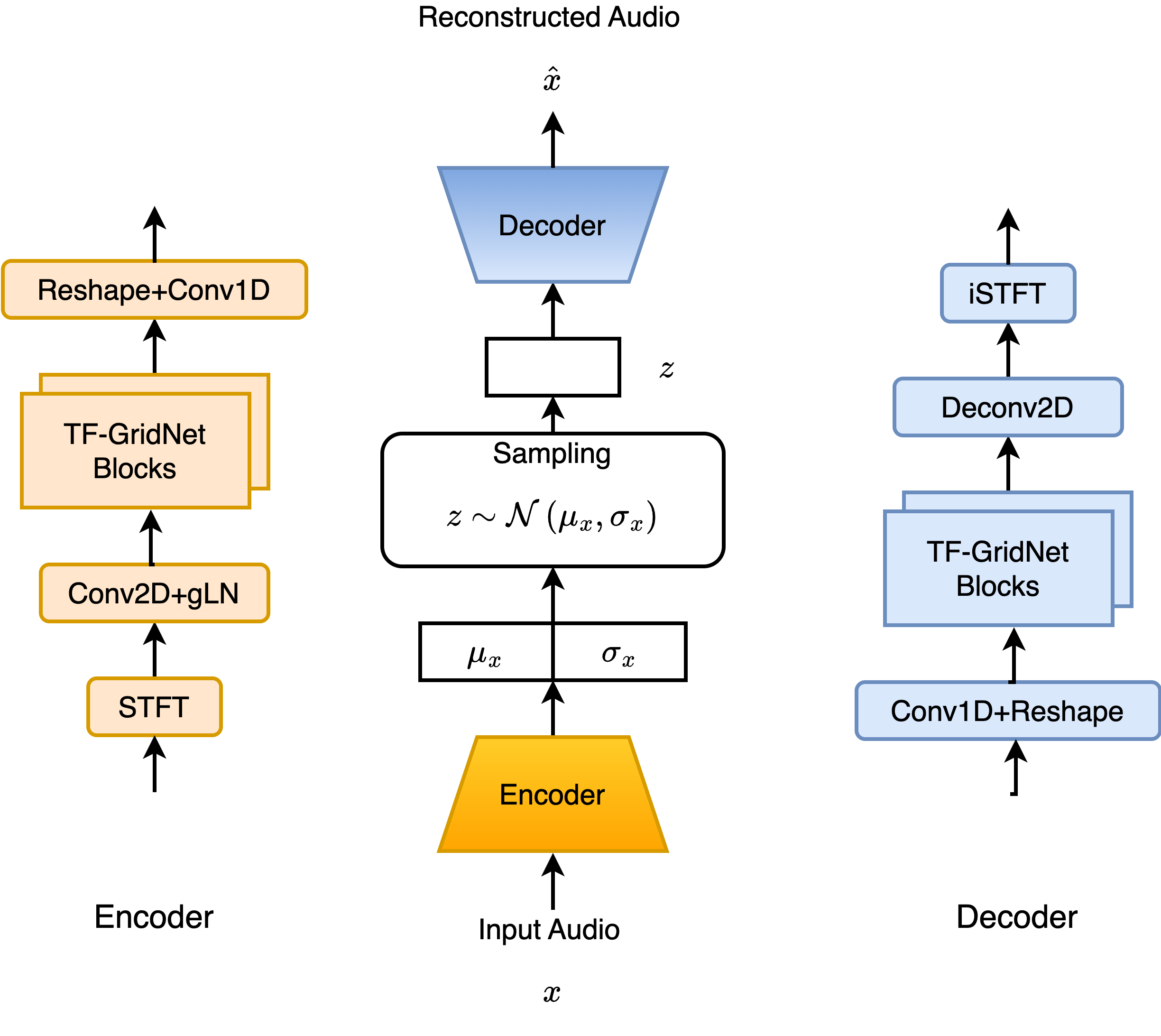}
\caption{The audio compressor architecture with encoder (orange) and decoder (blue) in detail.} 
\label{fig:compressor}
\end{figure}

\section{DiT-Flow Speech Enhancement}

In this section, we introduce DiT-Flow, a flow matching-based speech enhancement model built on the DiT architecture, designed to achieve multi-condition robustness, addressing challenges such as noise, reverberation, and compression.

\subsection{Overall pipeline}

Let $\boldsymbol{x}_d \in \mathbb{R}^{1 \times T}$ and $\boldsymbol{x}_c \in \mathbb{R}^{1 \times T}$ denote the distorted speech and the clean speech, respectively, where $T$ represent the audio length, in samples. The task of SE is to estimate the denoised speech signal $\boldsymbol{\hat{x}} \in \mathbb{R}^{1 \times T}$ from $\boldsymbol{x}_d$. As described in Section~\ref{sec:Audio compressor}, an audio compressor encodes the distorted into compact latent representations: $\boldsymbol{z}_d \in \mathbb{R}^{D \times L}$ for distorted speech, where $L$ denotes the number of frames in the latent space and $D$ is the dimensionality of each frame's feature vector.
A continuous transformation that maps the distribution of distorted speech to that of clean speech in the latent space is then learned by the flow-matching generative module. At inference time, as detailed in Section~\ref{sec:Target extractor}, the flow-matching module infers the latent target features $\boldsymbol{z}_{\hat{x}} \in \mathbb{R}^{D \times L}$ by solving the ODE in~\eqref{eq:ode}.
Finally, the decompressor reconstructs the estimated clean waveform $\boldsymbol{\hat{x}} \in \mathbb{R}^{1 \times T}$ from the predicted latent representation.

\subsection{Audio compressor}
\label{sec:Audio compressor}

The audio compressor is designed to transform raw waveforms into compact sequences of latent features. State-of-the-art approaches typically adopt time-domain variational autoencoders (VAEs) composed of multiple convolutional blocks~\cite{kumar2023high, hai2024ezaudio}. Inspired by recent advances in time–frequency (T-F) modeling for mixture signals~\cite{wang2023tf, li2024spmamba}, the compressor architecture from~\cite{wang2025solospeech, wang2025soloaudio} is adapted for the speech enhancement task.
As illustrated in Figure~\ref{fig:compressor}, the process begins by applying the Short-Time Fourier Transform (STFT) to the input waveform $\boldsymbol{x} \in \mathbb{R}^{1 \times N}$, yielding a complex-valued spectrogram $\boldsymbol{S} \in \mathbb{R}^{2 \times F \times \frac{N}{h}}$, where $N$ denotes the total number of audio samples, $h$ is the hop size, and $F$ is the number of frequency bins. The real and imaginary components of the STFT output are concatenated along the channel dimension to form the input representation for subsequent encoding.

TF-GridNet is employed as the backbone of our VAE~\cite{wang2023tf}, comprising stacked TF-GridNet blocks. Through the encoder, a latent tensor with shape $\mathbb{R}^{1 \times 2D \times \frac{N}{h}}$ is obtained, subsequently partitioned into a mean $\mu_x \in \mathbb{R}^{1 \times D \times \frac{N}{h}}$ and a variance $\sigma_x \in \mathbb{R}^{1 \times D \times \frac{N}{h}}$. A latent sample $z \sim \mathcal{N}(\mu_x, \sigma_x)$ is then drawn, yielding $z \in \mathbb{R}^{1 \times D \times \frac{N}{h}}$. The decoder symmetrically mirrors the encoder and reconstructs the waveform via the inverse STFT. Training proceeds in a combined generative–adversarial fashion~\cite{evans2024stableaudioopen} using three objectives: (1) a perceptually weighted multi-resolution STFT loss~\cite{steinmetz2020auraloss}; (2) an adversarial feature-matching loss with five convolutional discriminators as in Encodec~\cite{defossez2022high}; and (3) a Kullback–Leibler divergence penalty.

\begin{figure*}
  \centering
\includegraphics[width=17cm]{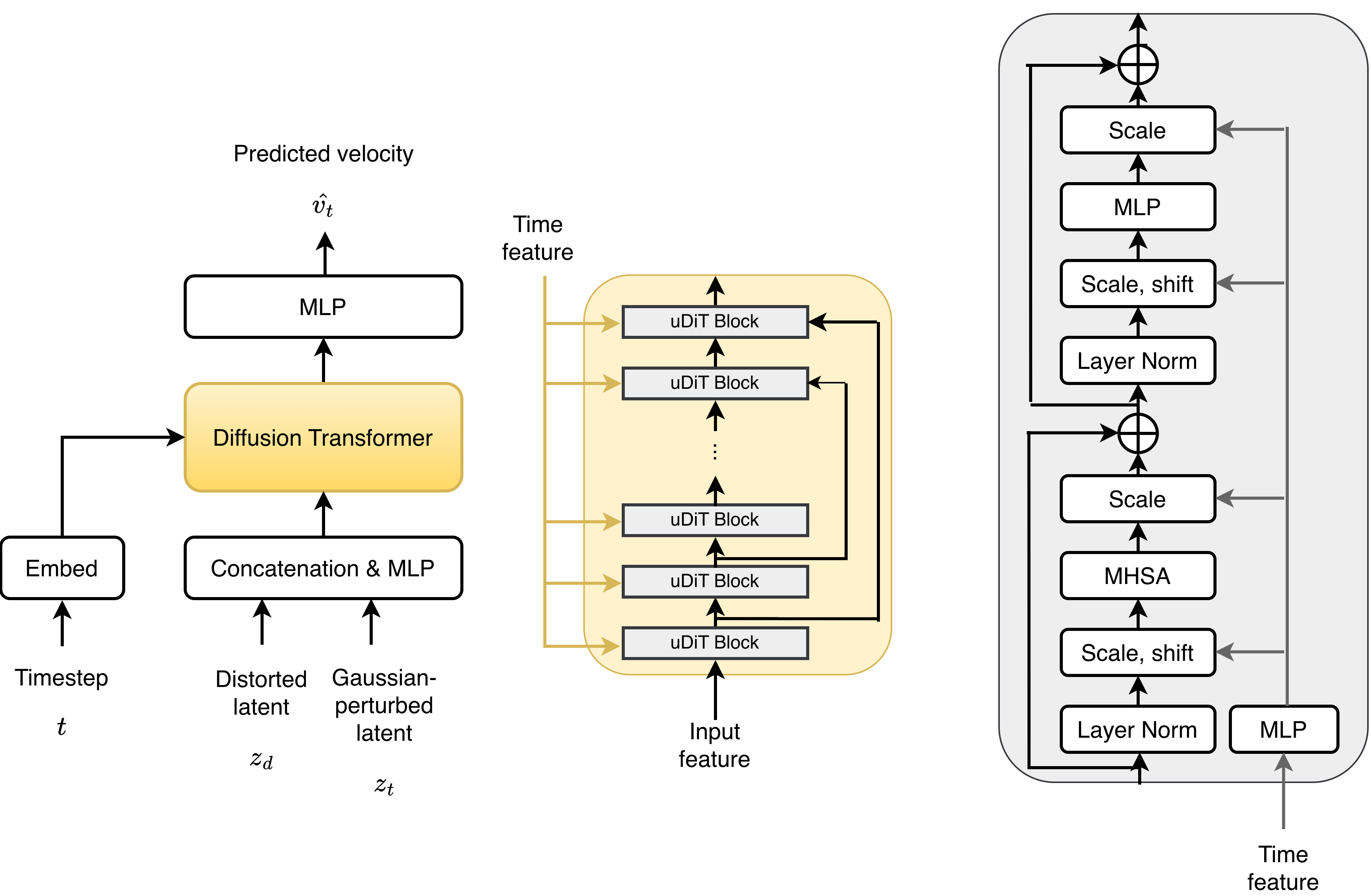}
\caption{The model backbones of the target extractor with Diffusion Transformer backbone (yellow) and uDiT block (grey).} 
\label{fig:extractor}
\end{figure*}

\begin{figure*}
  \centering
\includegraphics[width=17cm]{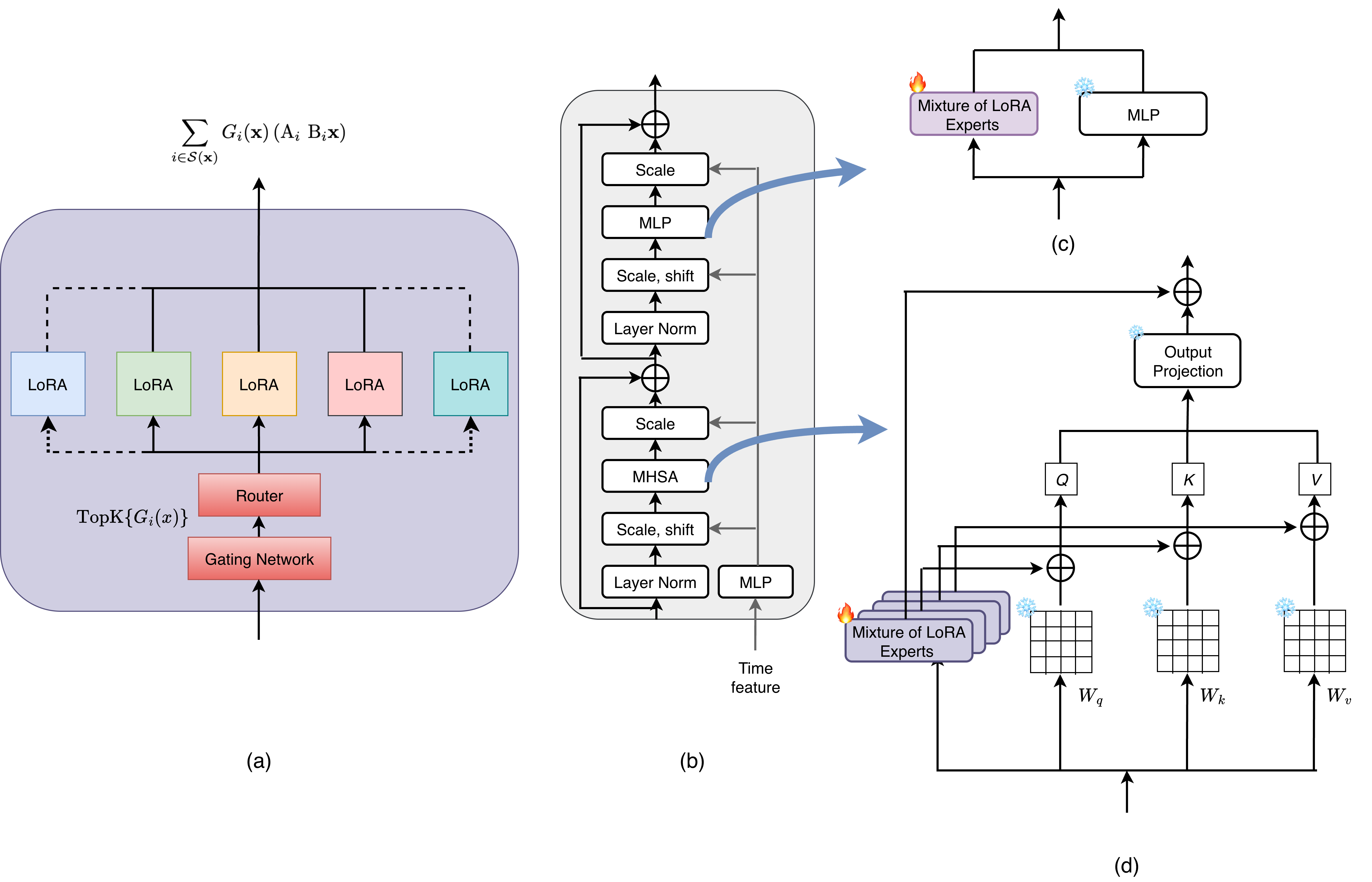}
\caption{Diagram of the Mixture of LoRA Experts in DiT-Flow model for data adaptation. (a) illustrates the basic Mixture of LoRA Experts mechanism. (b) In a UDiT block, replace the standard adaptation path in both the MHSA and MLP sublayers with Mixture of LoRA Experts modules (blue arrows), while keeping the original normalization and modulation structure. (c) The MLP modification. (d) The MHSA modification.} 
\label{fig:Mixture of LoRA Experts}
\end{figure*}

\subsection{Flow Matching Module}
\label{sec:Target extractor}

DiT-Flow is a flow-matching-based speech enhancement model that operates entirely within the latent space of a variational autoencoder (VAE), aiming to recover the target latent representation $\boldsymbol{z}_{\hat{x}}$. As illustrated in Figure~\ref{fig:extractor}, the core of the flow-matching module is a diffusion transformer equipped with extended skip connections, known as uDiT\cite{wang2025soloaudio}, a variant of the DiT architecture\cite{peebles2023scalable}.
The skip connections in uDiT bridge shallow and deep transformer blocks, allowing low-level features to bypass deeper layers. This architectural design facilitates more effective gradient flow and improves the training stability of the velocity prediction network.

During training, the distorted latent representation $\boldsymbol{z}_d$ is concatenated with its Gaussian-perturbed noisy latent $\boldsymbol{z}_t$ and, together with the flow time step $t$, fed into the uDiT backbone. An ordinary differential equation (ODE) solver is then employed to transform this sample from the base distribution to the target distribution, effectively recovering the enhanced latent representation.

\subsection{Extension of Mixture of LoRA Experts for domain adaptation}
\label{sec:ExtensionforDA}

Previous studies, e.g., \cite{zoph2022st}, suggested that it can significantly improve performance by fine-tuning the attention layer. Particularly, by introducing a mixture-of-experts mechanism within each MHSA block in uDiT shown in Figure~\ref{fig:Mixture of LoRA Experts}. To fully leverage the potential of DiT-Flow in robustness to multiple conditions, particularly for unseen distortions during training, we employ a Mixture of LoRA Experts fine-tuning structure by the modification of the FeedForward Network (FFN), i.e., MLP, and Multi-Head Self-Attention (MHSA). Figure~\ref{fig:Mixture of LoRA Experts} (a) illustrates the basic Mixture of LoRA Experts mechanism. A gating network produces input-dependent routing scores, and a router selects a small active set of LoRA experts $\mathcal{S}(x)$. The adaptation is then formed as a weighted combination of low-rank updates, $\sum_{i \in \mathcal{S}(x)} G_i(x)\left(A_i B_i x\right)$, so the backbone computation is preserved while the update is sparse and sample-specific. In Figure~\ref{fig:Mixture of LoRA Experts}(b), we apply this idea to a UDiT block for data adaptation, replacing the standard adaptation path in both the MHSA and MLP sublayers with Mixture of LoRA Experts modules (blue arrows), while keeping the original normalization and modulation structure (e.g., layer norm and scale/shift conditioning from the time feature) unchanged. Only the router, gating, and LoRA parameters are learned, and the pretrained block weights remain fixed. Figure~\ref{fig:Mixture of LoRA Experts} (c) zooms into the MLP modification, where the feed-forward transformation is augmented by a Mixture of LoRA Experts LoRA branch, enabling the block to switch among multiple low-rank experts to better match different data conditions without expanding the full MLP weights. Figure~\ref{fig:Mixture of LoRA Experts} (d) shows the analogous change for MHSA, where Mixture of LoRA Experts are attached to the attention projections (e.g., $W_q, W_k, W_v$, and output projection), allowing the model to adapt how it forms queries, keys, values as well as the projection in a routed, input-conditioned manner, and effectively tailoring attention behavior to the target domain while retaining the efficiency and stability of low-rank adaptation.

Besides, MoELoRA enables efficient cross-domain adaptation by leveraging its modular expert structure, whereby a pretrained model can be extended to new data domains through the addition of new LoRA experts. In speech enhancement, the intuition is natural, different experts can specialize in distinct acoustic aspects, e.g., noise families, reverberation profiles, device characteristics, or codec artifact patterns, and a router can select or combine experts based on the current input.

\begin{figure}
  \centering
\includegraphics[width=7cm]{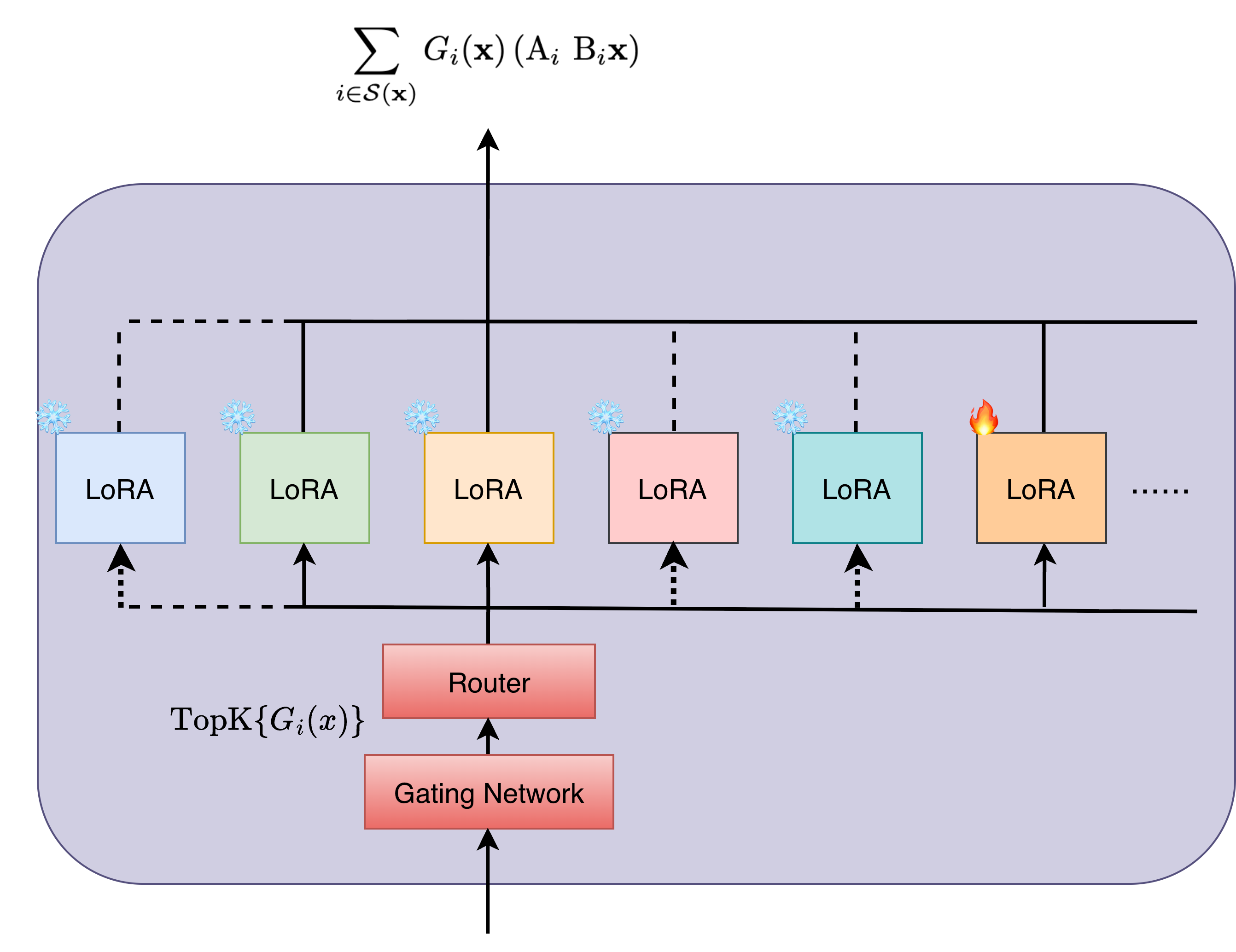}
\caption{The extentions of MoELoRA module with only training one single expert.} 
\label{fig:extending MoELoRA module}
\vspace{-2mm}
\end{figure}

The Figure~\ref{fig:extending MoELoRA module} illustrates how MoELoRA supports domain extension through additive LoRA experts. To adapt a pretrained model to a new domain, e.g., previously unseen distortion conditions in speech enhancement, a new LoRA expert is appended in parallel to the existing ones. During adaptation, training is restricted to the newly introduced expert and the router parameters, while the original experts are kept fixed, indicated by the snowflake icons. This isolates domain-specific learning to a small parameter set, preserving prior performance while making the update lightweight, efficient, and practical for rapid deployment.

\section{Experimental Setup}

\subsection{Dataset}
\label{sec:dataset}

\subsubsection{\textbf{StillSonicSet}}
To train and evaluate the performance of DiT-Flow, we conducted experiments on two versions of the StillSonicSet dataset: one containing only reverberant speech, and another augmented with noise and compression distortions (reverberant + noise + compression). For training, we randomly selected 50,000 utterances (approximately 90 hours of audio) from the training set. The full validation and test sets, each comprising 8 hours of speech, were used for model validation and final evaluation, respectively.

\subsubsection{\textbf{WSJ0+Reverb}}

The WSJ0+Reverb dataset is generated using clean speech data from the WSJ0 dataset and convolving each utterance with a simulated RIR by the pyroomacoustics engine \cite{scheibler2018pyroomacoustics}, which was used in \cite{richter2023speech, lemercier2023storm}. 

% URGENT dataset
\subsubsection{\textbf{URGENT}}

URGENT dataset \cite{li2026icassp} was originally employed in the URGENT challenge (Universal, Robust and Generalizable speech EnhancemeNT), which aims to build universal speech enhancement models for unifying speech processing in a wide variety of conditions\footnote{\url{https://urgent-challenge.github.io/urgent2026/}}. URGENT dataset mixtures are simulated from three categories of sources: speech, noise, and room impulse responses (RIRs). The dataset draws speech from a broad range of public speech corpora with different conditions and speaking styles, e.g., LibriVox, LibriTTS, VCTK, EARS, Multilingual Librispeech (MLS), CommonVoice 19.0, NNCES, SeniorTalk, VocalSet, ESD. The noise corpus is established by collecting noise clips from the AudioSet and FreeSound in DNS5 Challenge, WHAM!, FSD50K, Free Music Archive, plus simulated wind noise by the wind-noise simulator provided in \cite{lemercier2023wind} and RIRs from DNS5 / OpenSLR SLR28. This dataset considers the following seven distortions: (1) additive noise, (2) reverberation, (3) clipping, (4) bandwidth limitation, (5) codec loss (MP3 and OGG), (6) packet loss, and (7) wind noise. 

\subsubsection{\textbf{LibriCSS}}
Two more real recorded datasets were employed to validate the acoustic simulation gap between synthetic datasets and real data. The LibriCSS dataset is a real-recorded corpus derived from LibriSpeech, where utterances are concatenated to simulate conversations and replayed for capture with far-field microphones in a real room environment, rather than via simulation \cite{chen2020continuous}. Although LibriCSS was originally designed for speech separation, only the two recording conditions with a 0 overlap ratio (out of six total conditions) were selected. Using the provided Python scripts, 1416 utterances were generated for evaluation. 

\subsubsection{\textbf{RealMAN}}
To make the evaluation more challenging, RealMAN, a new relatively large-scale Real-recorded and annotated Microphone Array speech and Noise (RealMAN) dataset, was also used for testing, which includes 31 scenes(indoor, semi-outdoor, outdoor, transportation categories) \cite{yang2024realman}. Nine relevant scenes were chosen: Classroom2-3, LivingRoom2, 4-5, OfficeRoom2-4 and OfficeLobby, with 1692 utterances in total.

\subsection{Model configurations}

\subsubsection{\textbf{Audio compressor}}
In Audio compressor, we employed STFT with a window size of 40ms and hop size of 20ms, yielding latent representations at a temporal resolution of 50 Hz. The latent dimensionality is fixed at $D=128$. Within the encoder, a $3 \times 3$ complex-valued Conv2D layer (zero padding $1 \times 128$ channels) is followed by Group Normalization, after which three TF-GridNet blocks are applied. Each time-frequency unit uses a 128-dimensional embedding. Both the Unfold and Deconv1D operators adopt a kernel size and stride of 1, and each bidirectional LSTM contains 256 hidden units per direction. The self-attention module generates query and key tensors via $1 \times 1$ Conv2D layers with 512 output channels and employs 4 attention heads. Subsequent feature maps are reshaped and processed by a 128-channel Conv1D layer; all nonlinearities use the PReLU activation. The decoder mirrors the encoder configuration as mentioned in \ref{sec:Audio compressor}.

As mentioned, training proceeds in a combined generative–adversarial fashion~\cite{evans2024stableaudioopen} using three objectives: (1) Reconstruction Loss: We adopted a perceptually weighted, multi-resolution STFT loss with window lengths of [1280, 640, 320, 160, 80, 40, 20] and hop sizes of [320, 160, 80, 40, 20, 10, 5], respectively; (2) Adversarial Loss with Feature Matching: we utilizes a fixed mel bin size of 64, window sizes of [1280,640,320,160,80], and hop sizes of [320,160,80,40,20], following five convolutional discriminators as described in Encodec; (3) KLDivergence Loss: Down-weighted for KLDivergence Loss was set as $1\times10^{-4}$. The audio compressor consisted of a total of 49.3 million parameters. 

\subsubsection{\textbf{Flow-matching module}}
For the flow-matching module, the transformer backbone was configured with the following hyperparameters: 12 transformer layers, an embedding dimension of 384, and 6 attention heads. The model was trained using AdamW as the optimizer and a learning rate of $2 \times 10^{-4}$. The target extractor consists of approximately 50.6 million parameters. During inference, the number of ODE solver steps was set to 50.

\subsubsection{\textbf{Mixture-of-LoRA-Experts}}

Mixture of LoRA Experts extends the conventional single-LoRA adaptation scheme by integrating a mixture-of-experts mechanism into each self-attention block. Specifically, each block is augmented with a set of LoRA-based experts, implemented at the low rank $r=8$, together with a learned gating router. Model capacity and specialization are controlled by the experts with a number of 5. During forward propagation, a sparse gating strategy selects the top $ k$ experts, where $k=3$.

\subsection{Baseline methods}

To evaluate the efficiency of the proposed approach, we compared DiT-Flow with existing methods. Two diffusion-based models, i.e., SGMSE~\cite{richter2023speech} and StoRM~\cite{lemercier2023storm} were selected as the baselines. We trained those models on the same datasets mentioned in \ref{sec:dataset} using the authors’ official settings.

\subsection{Evaluation Metrics}

Standard metrics like PESQ~\cite{rix2001perceptual} and ESTOI are commonly used to assess speech enhancement models, but they may not be suitable for evaluating generative models~\cite{pirklbauer2023evaluation, fu2019learning}. This is because these metrics assume waveform alignment between the reference and enhanced signals, which generative models often disrupt due to minor misalignments or structural changes. \cite{kumar2025prose} shows diffusion baselines generating waveform details that are misaligned with the original speech, which can degrade alignment-sensitive intrusive metrics.

Log-Spectral Distance (LSD) was also adopted to evaluate the quality of enhanced speech signals, which measures the dissimilarity between the log-spectral representations of the clean and enhanced (or processed) speech signals. Lower LSD values indicate that the enhanced signal's spectral characteristics are closer to the clean signal's, suggesting better speech quality and less distortion.

To evaluate the perceptual quality of enhanced speech, we adopt DNSMOS P.835~\cite{reddy2021dnsmos}, a non-intrusive, neural-network-based metric developed by Microsoft. Unlike traditional intrusive metrics such as PESQ and STOI, DNSMOS P.835 does not require access to clean reference signals. It predicts three separate quality dimensions inspired by the ITU-T P.835 subjective evaluation protocol: SIG (speech quality), which measures the naturalness and clarity of the speech; BAK (background intrusiveness), which assesses how distracting the background noise is; and OVRL (overall quality), which reflects the combined perceptual impression of both speech and noise. These scores range from 1.0 (poor) to 5.0 (excellent) and closely approximate human mean opinion scores (MOS). This makes DNSMOS P.835 particularly suitable for benchmarking real-world speech enhancement models under diverse and challenging acoustic conditions.

Besides, speaker similarity (SIM) was further evaluated by computing the cosine similarity between the enhanced waveform and its clean reference using a pretrained WavLM-based speaker-verification model\footnote{at: \url{https://huggingface.co/microsoft/ wavlm-base-plus-sv}}.

\begin{table*}[!t]
\centering
\caption{Comparison of different systems trained on StillSonicSet with the mixture of reverberant, noise and codec-compression distortions and evaluated on mixture of reverberant, noise and codec-compression condition.}
\label{tab:dnsmos_rnc}
\resizebox{\textwidth}{!}{
\begin{tabular}{@{}llcccccccc@{}}
\toprule
\multirow{2}{*}{System} & \multirow{2}{*}{Model Type} 
& \multicolumn{7}{c}{Reverb+Noise+Codec-Compression} \\
\cmidrule(lr){3-9}  
& & PESQ $\uparrow$ & ESTOI $\uparrow$ & LSD $\downarrow$ & SIG $\uparrow$ & BAK $\uparrow$ & OVRL $\uparrow$ & Spk Sim $\uparrow$ \\
\midrule
Noisy (lower bound) & -- & 1.126 & 0.312 & 8.293 & 1.545 & 1.494 & 1.277 & 0.779  \\
SGMSE~\cite{richter2023speech} & Diffusion & 1.353 & 0.351 & 7.281 & 3.115 & 3.833 & 2.737 & 0.870 \\
StoRM~\cite{lemercier2023storm} & Diffusion & 1.302 & 0.431 & 5.413 & 2.996 & \textbf{3.969} & 2.601 & 0.837 
\\
Dit-Flow & Flow Matching &\textbf{1.389} & \textbf{0.458} & \textbf{4.506} & \textbf{3.301} & 3.723 & \textbf{2.906} & \textbf{0.880} 
\\

\bottomrule
\end{tabular}
}
\end{table*}

\begin{table*}[htbp]
\centering
\caption{Comparison of different systems trained on StillSonicSet with the mixture of reverberant, noise and codec-compression distortion and evaluated on reverberant-only condition.}
\label{tab:dnsmos_Reverb}
\resizebox{\textwidth}{!}{
\begin{tabular}{@{}llcccccccc@{}}
\toprule
\multirow{2}{*}{System} & \multirow{2}{*}{Model Type} 
& \multicolumn{7}{c}{Reverb only} \\
\cmidrule(lr){3-9}  
& & PESQ $\uparrow$ & ESTOI $\uparrow$ & LSD $\downarrow$ & SIG $\uparrow$ & BAK $\uparrow$ & OVRL $\uparrow$ & Spk Sim $\uparrow$ \\
\midrule
Noisy (lower bound) & -- & 1.324 & 0.497 & 3.383 & 2.080 & 2.386 & 1.684 & 0.895
\\
SGMSE~\cite{richter2023speech} & Diffusion & \textbf{2.011} & \textbf{0.632} & 4.031 & 3.166 & 3.882 & 2.775 & \textbf{0.943} 
\\
StoRM~\cite{lemercier2023storm} & Diffusion & 1.711 & 0.561 & 4.353 & 3.018 & \textbf{3.909} & 2.626 & 0.922 
\\
Dit-Flow & Flow Matching & 1.599 & 0.578 & \textbf{3.979} & \textbf{3.240} & 3.826 & \textbf{2.851} & 0.935 
\\

\bottomrule
\end{tabular}
}
\end{table*}

\begin{table*}[htbp]
\centering
\caption{Comparison of different systems trained on StillSonicSet with the mixture of reverberant, noise and codec-compression distortion and evaluated on noise-only condition.}
\label{tab:dnsmos_Noise}
\resizebox{\textwidth}{!}{
\begin{tabular}{@{}llcccccccc@{}}
\toprule
\multirow{2}{*}{System} & \multirow{2}{*}{Model Type} 
& \multicolumn{7}{c}{Noise only} \\
\cmidrule(lr){3-9}  
& & PESQ $\uparrow$ & ESTOI $\uparrow$ & LSD $\downarrow$ & SIG $\uparrow$ & BAK $\uparrow$ & OVRL $\uparrow$ & Spk Sim $\uparrow$ \\
\midrule
Noisy (lower bound) & -- & 1.203 & 0.696 & 3.809 & 2.840 & 2.125 & 2.008 & 0.938
\\
SGMSE~\cite{richter2023speech} & Diffusion & 1.314 & 0.497 & 6.158 & 3.444 & 3.519 & 2.949 & 0.895 
\\
StoRM~\cite{lemercier2023storm} & Diffusion & \textbf{1.585} & \textbf{0.648} & 8.522 & 3.418 & \textbf{4.026} & 3.105 & 0.930 
\\
Dit-Flow & Flow Matching & 1.575 & 0.612 & \textbf{6.001} & \textbf{3.461} & 3.916 & \textbf{3.174} & \textbf{0.936} 
\\

\bottomrule
\end{tabular}
}
\end{table*}

\begin{table*}[!t]
\centering
\caption{Comparison of different systems trained on StillSonicSet with the mixture of reverberant, noise and codec-compression distortions and evaluated on the mixture of reverberant+noise condition.}
\label{tab:dnsmos_Reverb+Noise}
\resizebox{\textwidth}{!}{
\begin{tabular}{@{}llcccccccc@{}}
\toprule
\multirow{2}{*}{System} & \multirow{2}{*}{Model Type} 
& \multicolumn{7}{c}{Reverb+Noise} \\
\cmidrule(lr){3-9}  
& & PESQ $\uparrow$ & ESTOI $\uparrow$ & LSD $\downarrow$ & SIG $\uparrow$ & BAK $\uparrow$ & OVRL $\uparrow$ & Spk Sim $\uparrow$ \\
\midrule
Noisy (lower bound) & -- & 1.124 & 0.450 & 4.604 & 1.788 & 1.543 & 1.376 & 0.831
\\
SGMSE~\cite{richter2023speech} & Diffusion & 1.248 & 0.373 & \textbf{7.223} & \textbf{3.318} & 3.234 & 2.719 & 0.865
\\
StoRM~\cite{lemercier2023storm} & Diffusion & \textbf{1.464} & 0.469 & 8.758 & 3.174 & 3.823 & 2.792 & \textbf{0.869} 
\\
Dit-Flow & Flow Matching & 1.378 & \textbf{0.484} & 7.576 & 3.189 & \textbf{3.856} & \textbf{2.818} & 0.858
\\

\bottomrule
\end{tabular}
}
\end{table*}

\section{Results}

\subsection{Comparison to baselines}
Table \ref{tab:dnsmos_rnc} - Table \ref{tab:dnsmos_Reverb+Noise} compare DiT-Flow with baselines and summarize the performance of different systems under several different conditions. When it comes to the metrics, naturalness is a critical factor in speech enhancement, as it reflects how close the enhanced speech sounds to natural human speech. The DNSMOS evaluation consists of three main metrics: SIG (Signal Quality), BAK (Background Artifacts) and OVRL (Overall Quality). Speaker cosine similarity (Spk Sim) was computed to evaluate the speaker identity preservation. Besides, several traditional metrics, e.g., PESQ, ESTOI, and LSD, are listed to demonstrate the performance of different systems on perceptual metrics.

It should be noted that the first row in each table reports metrics computed directly on the distorted input signals, which serve as a lower bound. As expected, when applying single type of distortion to each utterance, e.g., Table \ref{tab:dnsmos_Reverb} and Table \ref{tab:dnsmos_Noise}, metrics computed directly on the distorted input signals show better results compared with those involving multiple distortions, e.g., Table \ref{tab:dnsmos_rnc} and Table \ref{tab:dnsmos_Reverb+Noise}, indicating the increase of complexity level with more types of distortions mixed into one utterance.

Considering the real-world teleconferencing situation, a more realistic model is to convolve speech and any in-room noise with their RIRs, sum them, and then encode with codec. To emulate the actual scenarios, we specifically evaluate the compression effect after pre-processed with other distortion types, i.e., noise and reverberation, instead of solely introducing Codec-compression to clean speech. Therefore, four different conditions: Reverb-only, Noise-only, Reverb+Noise and Reverb+Noise+Codec-Compression, are selected to compare the performance of different speech enhancement systems.

As shown in Table~\ref{tab:dnsmos_rnc} to Table~\ref{tab:dnsmos_Reverb+Noise}, performance varies across models depending on the evaluation metric. Our proposed DiT-Flow model achieves the highest SIG score across Reverb-only, Noise only and Reverb+Noise+Codec-Compression conditions, suggesting that it preserves speech quality better than the other models. It also shows the highest OVRL scores across both conditions, indicating a good balance of signal quality and background noise. Notably, under the more challenging Reverb+Noise+Compression condition, DiT-Flow outperforms all competing models, indicating strong robustness to multiple distortions. For background intrusiveness (BAK), StoRM provides the strongest noise suppression, attaining the highest scores across both conditions. However, this comes at the cost of slightly lower overall quality (OVRL) and speech quality (SIG), reflecting a trade-off between noise reduction and naturalness. Additionally, Spk Sim scores from DiT-Flow are competitive, reflecting the model's ability to maintain the speaker's identity.

\begin{table*}[!t]
\centering
\caption{Comparisons on real-recorded datasets. The performance of different speech enhancement systems are compared by using real-recorded datasets: LibriCSS and RealMAN. For both two real-recorded datasets, all three speech enhancements obtain a better performance when trained on StillSonicSet, especially on more complex dataset, realMAN. }
\label{tab:Comparison_real_recorded}
\resizebox{\textwidth}{!}{
\begin{tabular}{@{}llccccccccc@{}}
\toprule
\multirow{2}{*}{System} & \multirow{2}{*}{Trained datatset} & \multirow{2}{*}{RTF$\downarrow$}
& \multicolumn{4}{c}{LibriCSS \cite{chen2020continuous}} 
& \multicolumn{4}{c}{RealMAN \cite{yang2024realman}}  \\
\cmidrule(lr){4-7} \cmidrule(lr){8-11} 
& & & SIG $\uparrow$ & BAK $\uparrow$ & OVRL $\uparrow$ & Spk Sim $\uparrow$ 
  & SIG $\uparrow$ & BAK $\uparrow$ & OVRL $\uparrow$ & Spk Sim $\uparrow$ \\
\midrule
Noisy (lower bound) & N/A & N/A & 2.756 & 3.521 & 2.209 & 0.895 
                  & 1.970 & 2.204 & 1.417 & 0.860 \\
\midrule
SGMSE~\cite{richter2023speech} & WSJ0+Reverb \cite{lemercier2023storm} & \multirow{2}{*}{0.565} & 2.854 & 3.830 & 2.363 & 0.952 
                  & 2.473 & 3.031 & 1.863 & 0.892 \\
SGMSE~\cite{richter2023speech} & StillSonicSet & & 2.955 & 3.901 & 2.479 & 0.951 
                  & 2.829 & 3.623 & 2.324 & 0.883 \\
\midrule
StoRM~\cite{lemercier2023storm} & WSJ0+Reverb \cite{lemercier2023storm}& \multirow{2}{*}{0.494} & 2.961 & 3.822 & 2.485 & \textbf{0.955} & 2.481 & 2.976 & 1.884 & 0.891 \\
StoRM~\cite{lemercier2023storm} & StillSonicSet & & \textbf{2.963} & 3.946 & 2.490 & 0.953 
                  & 2.661 & 3.662 & 2.154 & 0.886\\
\midrule
DiT-Flow & WSJ0+Reverb \cite{lemercier2023storm}& \multirow{2}{*}{0.230} & 2.938 & 3.902 & 2.476 & 0.917 
              & 2.754 & 3.413 & 2.184 & 0.885  \\
DiT-Flow & StillSonicSet &  & 2.935 & \textbf{3.950} & \textbf{2.503} & 0.928 
              & \textbf{2.911} & \textbf{3.684} & \textbf{2.402} & \textbf{0.896} \\

\bottomrule
\end{tabular}
}
\end{table*}

When evaluating performance using conventional objective metrics such as PESQ\cite{rix2001perceptual} and ESTOI, it is important to note that these metrics may not be well-suited for assessing generative models~\cite{pirklbauer2023evaluation, fu2019learning} due to the fact that those metrics are reference-based metrics with an explicit alignment stage, and time-structure changes matter. Additionally, compared to existing synthetic datasets that often assume idealized conditions, e.g., empty, box-shaped rooms with simplified room impulse responses, the StillSonicSet offers a more acoustically complex and realistic simulation of real-world scenarios. As a result, none of the baseline models reach the performance levels reported in their original publications. When additional distortion factors, such as the mxiture of noise, reverb and compression are introduced, all models experience a moderate drop in performance. SGMSE performs best under the Reverb-only condition, achieving a PESQ score of 2.011, but declines to 1.35 in the Reverb + Noise + Compression condition. In contrast, DiT-Flow demonstrates consistently solid performance across both conditions, with PESQ scores of 1.599 (Reverb-only) and 1.389 (Reverb + Noise + Compression). While it does not achieve the highest PESQ score, DiT-Flow maintains a strong balance between naturalness and intelligibility. SGMSE also performs well on this metric, with scores of 0.632 and 0.351, respectively, reflecting good intelligibility but not at the level achieved by DiT-Flow.
Furthermore, DiT-Flow leads in terms of Log-Spectral Distance (LSD), with the lowest scores of 3.979 (Reverb-only) and 4.506 (Reverb + Noise + Compression), suggesting it introduces the least spectral distortion among the models. This highlights DiT-Flow’s ability to preserve the spectral characteristics of the original clean signal more effectively than its counterparts.

Overall, DiT-Flow outperforms all other models in Reverb+Noise+Compression conditions, showing a robust enhancement across multiple distortions.

\subsection{Generalization to unseen real-recorded data}

Table \ref{tab:Comparison_real_recorded} compares the performance of different speech enhancement systems using real-recorded datasets: LibriCSS and RealMAN as mentioned in Section \ref{sec:dataset}. Similarly, the first row in each table reports metrics computed directly on the distorted input signals as a lower bound. For both two real-recorded datasets, all three speech enhancements obtain a better performance when trained on StillSonicSet, especially on more complex dataset, realMAN. 

For LibriCSS dataset, DiT-Flow (WSJ0+Reverb) achieves the top SIG, showing good preservation of naturalness, while DiT-Flow (StillSonicSet) yields the best BAK and OVRL, confirming stronger noise suppression and overall perceptual quality. Besides, DiT-Flow (StillSonicSet) remains competitive Spk Sim score, balancing speech quality with identity preservation. On the other hand, RealMAN is a more challenging dataset due to Mandarin speech and cross-lingual robustness as well as diverse reverberant scenes, e.g., Classroom, LivingRoom, etc. DiT-Flow trained on StillSonicSet again achieves the best performance with the SIG, strongest BAK suppression, and the highest OVRL, demonstrating robustness across languages and environments. DiT-Flow also obtains the top Spk Sim, surpassing StoRM despite the cross-lingual setting. Models trained on WSJ0+Reverb show limited generalization, particularly on RealMAN, reflecting the gap between synthetic reverberant data and real acoustics. In contrast, StillSonicSet training consistently boosts DNSMOS scores, especially for BAK and OVRL, validating its design with realistic geometries and occlusions. 

Real-time Factor (RTF) is a performance metric measuring the efficiency of data processing systems, specifically defined as the ratio of processing time to the duration of the input data. Table \ref{tab:Comparison_real_recorded} demonstrates a significant improvement of DiT-Flow for the efficiency of data processing. Two diffusion-based models, SGMSE and StoRM, show relatively high real-time factors, which are more than twice of RTF for DiT-Flow.  This potentially fulfill the task at larger scale but lower computation cost and time.

Overall, DiT-Flow trained on StillSonicSet achieves the strongest balance, which improves perceptual quality (SIG, BAK, OVRL) while preserving speaker similarity across both English and Mandarin recordings, and highlights its multi-condition and cross-lingual robustness. StillSonicSet can be a better choice, and models trained on StillSonicSet achieve markedly stronger generalization to a more complex real-world condition.

\begin{table*}[htbp]
\centering
\caption{Comparison  of different data adaptation strategies for DiT-Flow under a challenging
distribution shift scenario.}
\label{tab:DA}
\resizebox{\textwidth}{!}{
\begin{tabular}{@{}llcccccccc@{}}
\toprule

\multirow{2}{*} {Model} & \multirow{2}{*} { Params (\%)} & \multirow{2}{*} {Finetuned dataset (h)} 
& \multicolumn{7}{c} {Metric}\\
\cmidrule(lr){4-10}
& & & SIG $\uparrow$ & BAK $\uparrow$ & OVRL $\uparrow$ & Spk Sim $\uparrow$ & PESQ~$\uparrow$ & ESTOI~$\uparrow$ & LSD~$\downarrow$
\\
\midrule
Distorted (lower bound) & N/A & N/A & 2.467 & 1.999 & 1.810 & 0.891 & 1.272 & 0.624 & 5.461 \\
\midrule
Pretrained DiT-Flow & 100 & N/A & 3.352 & 3.978 & 3.063 & 0.959 & 1.954 & 0.774 & 3.535\\
\midrule
\multirow{2}{*}{Trained from scratch} & \multirow{2}{*}{100} & 12 & 3.282 & 3.855 & 2.944 & 0.939 & 1.766 & 0.718 & 3.190\\
\cmidrule(lr){3-10}
& & 30 & 3.398 & 3.952 & 3.087 & 0.952 & 1.926 & 0.761 & 3.092\\

\midrule
\multirow{2}{*}{Full finetune} & \multirow{2}{*}{100} & 12 & 3.419 & 3.968 & 3.113 & 0.963 & 2.129 & 0.800 & 2.955\\
\cmidrule(lr){3-10}
& & 30 & 3.438 & 3.957 & 3.124 & 0.964 & \textbf{2.146} & 0.806  & \textbf{2.948}\\
\midrule
\multirow{2}{*}{LoRA} & \multirow{2}{*}{0.5} & 12 & 3.435 & 3.977 & 3.131 & 0.962 & 2.063 & 0.791 & 3.059\\
\cmidrule(lr){3-10}
& & 30 & 3.437 & 3.989 & 3.139 & 0.962 & 2.064 & 0.793 & 3.064\\
\midrule

\multirow{2}{*}{MoELoRA(MLP+Attn)} & \multirow{2}{*}{4.9} & 12 & 3.440 & 3.984 & 3.139 & 0.964 & 2.110 & 0.801 & 3.024\\
\cmidrule(lr){3-10}
& & 30 & \textbf{3.442} & \textbf{3.991} & \textbf{3.144} & \textbf{0.964} & 2.122 & \textbf{0.802} & 3.018\\

\bottomrule
\end{tabular}
}
\end{table*}

\subsection{Adaptation to unseen distortions}

Table \ref{tab:DA} presents a comprehensive comparison of different data adaptation strategies for DiT-Flow under a challenging distribution shift scenario. The pretrained DiT-Flow model is trained on approximately 180 hours of data (115122 pairs of utterances) from URGENT dataset containing only noise and reverberation distortions. In contrast, both the finetuning and evaluation datasets include seven distortion types, comprising not only noise and reverberation but also five unseen distortions: clipping, bandwidth limitation, codec loss (MP3 and OGG), packet loss, and wind noise. This setting allows us to systematically examine how different finetuning strategies adapt a pretrained model to heterogeneous and previously unseen acoustic conditions.

The first row reports metrics computed directly on the distorted input signals, which serve as a lower bound. As expected, all objective and perceptual scores are substantially degraded, with low SIG, BAK, OVRL, PESQ, and ESTOI values, and a high LSD. This confirms the severity of the distortions in the evaluation data and establishes a strong baseline for assessing enhancement performance.

The pretrained DiT-Flow model on URGENT, evaluated directly on unseen noisy data without finetuning, demonstrates a substantial improvement over the noisy baseline across all metrics. Notably, BAK increases from 1.999 to 3.978, indicating that large-scale pretraining enables the model to learn robust noise suppression priors. Improvements in SIG, OVRL, PESQ, and ESTOI further suggest that pretraining captures general speech and noise characteristics that transfer beyond the original training distortions. However, despite these gains, the pretrained model does not achieve optimal performance under the expanded distortion set. Metrics such as OVRL (3.063) and PESQ (1.954) remain noticeably below those achieved by finetuned models. This highlights an important limitation, that is, pretraining on a limited distortion space does not fully generalize to more diverse and non-stationary distortions, particularly those involving non-linear effects, e.g., clipping, or temporal corruption, e.g., packet loss.

Models trained from scratch on the same small finetuning datasets (12 or 30 hours) consistently underperform the pretrained model across all metrics. Even though these models are exposed to the same unseen distortions as the finetuned models, they fail to match the pretrained DiT-Flow in SIG, BAK, OVRL, PESQ, and ESTOI. This result shows the data inefficiency of training from scratch in low-resource settings and confirms the necessity of large-scale pretraining. Importantly, this comparison demonstrates that performance gains observed in finetuned models are not merely due to exposure to unseen distortions, but rather stem from adapting a strong pretrained representation, rather than relearning enhancement behavior from limited data.

Full finetuning of the pretrained DiT-Flow model, where all parameters are updated using the finetuning datasets, yields consistent improvements across nearly all metrics. In particular, full finetuning achieves the best PESQ (2.129) and lowest LSD (2.955), reflecting superior spectral fidelity and reduced distortion. OVRL and SIG are also improved relative to the pretrained model, while speaker similarity remains high. These results indicate that full finetuning is highly effective at adapting the model to complex, unseen distortions. However, this approach requires updating 100\% of the model parameters, which significantly increases computational cost and memory usage, and may raise concerns about overfitting or catastrophic forgetting in practical deployment scenarios.

LoRA finetuning offers a strong parameter-efficient alternative. With only a small fraction of trainable parameters, LoRA achieves performance comparable to full finetuning in terms of SIG and OVRL, while maintaining high speaker similarity. However, LoRA lags behind full finetuning in PESQ and LSD, suggesting limitations in modeling fine-grained spectral distortions. This gap can be attributed to the single-expert nature of LoRA, which constrains its ability to adapt flexibly to the diverse and heterogeneous distortion patterns present in the unseen dataset.

The strongest parameter-efficient results are obtained when MoELoRA is applied to both MLP (FFN) and attention layers. This configuration achieves the highest SIG, BAK, and OVRL among all parameter-efficient methods, and approaches or even matches full finetuning in ESTOI, while using less than 5\% of trainable parameters. These results indicate that extending expert adaptation to attention layers enables more effective modeling of temporal dependencies and long-range structure, which are particularly important for distortions such as packet loss, codec artifacts, and wind noise.

Taken together, these results demonstrate that while large-scale pretraining provides a strong foundation, explicit adaptation is essential for handling unseen and heterogeneous distortions. Full finetuning offers the best scores of PESQ and LSD but at high computational cost. In contrast, MoELoRA achieves a favorable balance between performance and efficiency by enabling expert specialization through mixture modeling and load balancing. Notably, MoELoRA with attention adaptation achieves the overall best performance across perceptual, intelligibility, and spectral distortion metrics, while preserving speaker similarity and updating only a small fraction of parameters. This highlights the effectiveness of mixture-based, parameter-efficient finetuning for robust speech enhancement under real-world distribution shifts.

Overall, DiT-Flow stands out as the most well-rounded model, achieving the best balance between signal quality, intelligibility, and spectral fidelity across various speech enhancement scenarios, which demonstrates its Multi-Condition Robustness.

\section{Conclusions}

In this work, we introduced DiT-Flow, a flow-matching-based framework for generalized speech enhancement, built upon the DiT backbone and trained to be robust across a wide range of acoustic distortions, including noise, reverberation, and codec compression. To address limitations in existing synthetic datasets, we also proposed StillSonicSet, a challenging dataset specifically designed for stationary sound sources in acoustically diverse environments. Constructed using the SonicSim toolkit and building upon the original SonicSet resources, StillSonicSet captures a broad spectrum of realistic conditions by incorporating complex room geometries, varied surface materials, and natural occlusions such as furniture and architectural structures. This marks a significant improvement over traditional datasets that rely on simplified, shoebox-style room impulse response (RIR) simulations. Through extensive training and evaluation on StillSonicSet, designed to better reflect real-world scenarios, DiT-Flow consistently outperformed baseline models, achieving the best balance among signal quality, intelligibility, and spectral fidelity. These results underscore DiT-Flow's strong multi-condition robustness and its effectiveness in handling diverse and challenging acoustic conditions. Despite ongoing efforts to expand synthetic data realism, a persistent bottleneck in SE is the inevitable mismatch between training and deployment conditions. By integrating LoRA with the MoE framework into DiT-Flow, the number of updated parameters is dramatically reduced during finetuning process, making data adaptation feasible under limited compute and latency budgets. We achieve both parameter-efficient and high-performance training for DiT-Flow robust to multiple distortions. Therefore, parameter-efficient adaptation offers a promising path toward robust SE in realistic pipelines, where training datasets cannot fully anticipate the distortions encountered at test time.

\bibliographystyle{IEEEtran}
\bibliography{refs}

\vfill

\end{document}